\documentclass[aps,pre,twocolumn,superscriptaddress,showpacs,floatfix]{revtex4}
\usepackage{graphicx}
\usepackage{amsmath}
\begin{document}
\title{Accelerated kinetic Monte Carlo algorithm 
for diffusion limited kinetics}
\author{V. I. Tokar}
\affiliation{Universit{\'e} Louis Pasteur, CNRS, IPCMS, 23 rue du Loess,
F-67034 Strasbourg, France}
\affiliation{Institute of Magnetism, National Academy of Sciences,
36-b Vernadsky st., 03142 Kiev-142, Ukraine}
\author{H. Dreyss\'e}
\affiliation{Universit{\'e} Louis Pasteur, CNRS, IPCMS, 23 rue du Loess,
F-67034 Strasbourg, France}
\date{\today}
\begin{abstract}
If a stochastic system during some periods of its evolution can be
divided into non-interacting parts, the kinetics of each part can be
simulated independently. We show that this can be used in the
development of efficient Monte Carlo algorithms. As an illustrative
example the simulation of irreversible growth of extended one
dimensional islands is considered. The new approach allowed to simulate
the systems characterized by parameters superior to those used
in previous simulations.
\end{abstract}
\maketitle
A unique feature of the kinetic Monte Carlo (kMC) technique which to a
large extent underlies its wide acceptance in physics
is its ability to provide essentially exact data describing complex
far-from-equilibrium phenomena \cite{binder}.  The technique, however,
is rather demanding on computational resources which in many cases
makes the simulations either impractical or altogether impossible
\cite{ratsch-venables-review,no-scaling}.  As was pointed out in
Ref.\ \cite{ratsch-venables-review}, the major cause of the low
efficiency of kMC is the large
disparity between the time scales of the participating processes.  In
fact, it is the fastest process which slows down the simulation the
most.  As a remedy it was suggested that the fast processes were
described in some averaged, mean-field manner.  These and similar
observations lie at the hart of various approximate multi-scale schemes
(see, e.\ g., Refs.\
\cite{ratsch-venables-review,multiscalePRL,1Dmultiscale,2Dmultiscale,%
3Dmultiscale}).

The approximate implementations, however, deprive kMC of its major
asset---the exactness.  As a consequence, it cannot serve as a
reliable tool for resolving controversial issues, such, e.\ g., as
those arising in connection with the scaling laws governing the
irreversible epitaxial growth (see Refs.\
\cite{no-scaling,1Dmultiscale,famarescu} and references therein).

Recently, an exact kMC scheme called by the authors the first-passage
algorithm (FPA) was proposed which avoids simulating all the hops of
freely diffusing atoms and using instead analytic solutions of an
appropriate diffusion equation \cite{FPT_MC}.  It is premature yet draw
definite conclusions about the efficiency of the algorithm tested only
on one system, at least before additional technical issues improving
its efficiency are published by the authors. However, the authors
themselves note that there are problems in the treatment of closely
spaced atoms.  This makes it difficult to use FPA in simulating the
diffusion limited kinetics in such cases when along with large empty
spaces where the analytic description is efficient there exist the
reaction zones where the particle concentrations are high as, e.\ g.,
in the vicinity of islands during the surface growth.  Furthermore,
because the majority of kMC simulations are performed with the use of
the by now classic event-based algorithm (EBA) of Ref.\ \cite{n-fold},
the FPA algorithm would be difficult to use in the upgrade of the
existing code.  This is because FPA is completely different from EBA
and its application would require a new code to be created from the
scratch.  In some cases this may be more time-consuming than the use of
the available EBA code.

The aim of the present paper is to propose an exact
accelerated kMC algorithm which extends the EBA in such a way that in
the case of the diffusion limited systems only the atoms which are
sufficiently well separated from the reaction zones are treated with
the use of exact diffusion equations while in the high-density regions
the conventional EBA is used.

The algorithm we are going to present can be applied to any separable
model.  For concreteness, we present it using as an example a simple
(but non-trivial---see \cite{amar_popescu} and references therein)
example of the irreversible growth in one dimension (1d)
\cite{1992,1Dscaling,amar_popescu}.  Its generalizations to other
systems are completely straightforward,

Our approach is based on the observation that the fastest process in
the surface growth is the hopping diffusion of the isolated atoms (or
monomers) \cite{ratsch-venables-review}.  Random walk on a lattice is
one of the best studied stochastic phenomena with a lot of exact
information available.  In cases when the monomers are well separated
from each other and from the growth regions, the analytical description
of their diffusion can be computationally much less demanding than
straightforward kMC simulation.

In the model of irreversible growth the atoms are deposited on the
surface at rate $F$ where they freely diffuse until meeting either
another atom or an island edge which results either in the nucleation
of a new island or in the growth of an existing one, respectively. To
illustrate the strength of our approach, we will study the limit of low
coverages $\theta\to0$ because in Ref.\ \cite{no-scaling} this limit
was considered to be difficult to simulate in the case of extended
islands.  Because the scaling limit corresponds to
\begin{equation} 
\label{ R}
R\equiv D/F\to\infty,
\end{equation}  
(where $D$ is the diffusion constant) i.\ e., to very low deposition
rates, and, furthermore, because the covered regions are also small due
to low $\theta$, we found it reasonable to neglect nucleation on the
tops of islands by assuming them to be monolayer-high.

In its simplest implementation our algorithm is based on a subdivision
of the monomers into two groups (A and B) which at a given moment are
considered to be active (A) and passive (B) ones with respect to the
growth processes.  The passive monomers are those which are too far
away from the places of attachment to existing islands or of nucleation
of new ones.  This can be quantified with the use of a separation
length $L$.  Thus, an atom is considered to be passive if it is
separated from a nearest island by more then $L$ sites or if its
separation from a nearest monomer exceeds $2L$.  The monomers which do
not satisfy these restrictions are considered to be actively
participating in the growth and thus belonging to the group A.  It is
the passive atoms B that we are going to treat within an analytical
approach instead of simulating them via kMC.  Thus, in contrast to FPA
where all atoms should be boxed, in our algorithm we may box only those
which will spend some appreciable time inside the boxes and will not
need to be quickly re-boxed as in the FPA algorithm with closely spaced
atoms.

Formally this is done as follows.  Let us place all B
atoms in the middle of 1d ``boxes'' of length $L_{\mbox{Box}}=2L+1$.
Assuming the central site has the coordinate $i=0$, the initial probability
distribution is of the Kronecker delta form
\begin{equation} \label{ init}
p(i,t=0)=\delta_{i0},
\end{equation}
where the time variable $t$ counts the time spent by the atom inside
the box.  With the atomic hopping rate set to unity, the evolution of
the probability distribution of an atom {\em inside} the box satisfies
the equations
\begin{subequations} 
\label{ group}
\begin{eqnarray}
\label{ conserving}
\frac{\partial p(i,t)}{\partial t}
&=&\frac{1}{2} p(i+1,t)+\frac{1}{2} p(i-1,t)-p(i,t)\\
\frac{\partial p(\pm L,t)}{\partial t}&=&\frac{1}{2}
p(\pm (L-1),t)-\frac{1}{2}p(\pm L,t),
\label{ conserving2}
\end{eqnarray}
\end{subequations} 
where $|i|<L$.  The first equation expresses the conservation of
probability on the interior sites $i\not=\pm L$.  The change of
probability on site $i$ given by the time derivative on the left hand
side comes from the probability of atoms hopping from neighbor sites
$i\pm1$ (two positive terms on the right hand side) minus the
probability for the atom to escape the site.  The ``in'' terms have
weights 1/2 because the atoms have two equivalent directions to hop.
The boundary equations (\ref{ conserving2}) differ only in that there
are neither incoming flux from the outside of the box, nor the outgoing
flux in this direction.

The solution at an arbitrary time can be written as
\begin{equation} 
\label{ p_t}
p(i,t)=L_{\mbox{Box}}^{-1}\left[1+2\sum_{m=1}^{L}
e^{-\epsilon_mt}\cos(\alpha mi)\right],
\end{equation} 
where 
\begin{equation} \label{ e_def}
\alpha=2\pi/L_{\mbox{Box}}\qquad
\mbox{and}\qquad\epsilon_m=2\sin^2(\alpha m/2).
\end{equation} 
The distribution Eq.\ (\ref{ p_t}) satisfies Eqs.\ (\ref{ group}) as
can be checked by direct substitution.  
The initial condition Eq.\ (\ref{ init}) as well as the probability
conservation $\sum_i p(i,t)=1$ can be verified with the use of
Eq.\ 1.342.2 from Ref.\ \cite{GR}.  In our algorithm we will need to
repeatedly calculate $p(i,t)$, so its efficient calculation is
important.  Eq.\ (\ref{ p_t}) is formally a discrete Fourier transform,
so it is natural to use an FFT algorithm.  Because our choice for the
position of the atom in the center of the box makes the box length odd
($L_{\mbox{Box}}=2L+1$), we used the radix-3 algorithm of
Ref.\ \cite{FFT}, so the sizes of all our boxes below are powers of 3.

The gain in the speed of the simulation is achieved because as long as
atoms B stay within the boxes we do not waste computational resources
to simulate them by knowing that they evolve according to Eq.\ (\ref{ p_t}).  

\begin{figure} 
\includegraphics[viewport = 0 0 386 250, scale = 0.6]{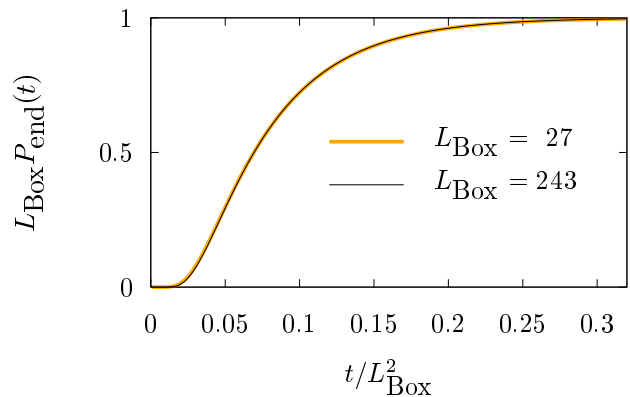}
\caption{\label{FIG1}(Color online) Time-dependent probability rate
$P_{\mbox{end}}(t)$ for the boxed atom to escape from the box.}
\end{figure}

Obviously, sooner or later the atomic configuration will change so that
the A-B division will cease to be valid.  This happens, in particular,
when an atom leaves the box.  Because the hopping in the model is
allowed only at the nearest neighbor (NN) distance, only the atoms at sites
$\pm L$ may leave the box.  With the hopping probability being 1/2 at
each side, the probability of an atom to leave the box is
\begin{equation} \label{ Pend}
P_{\mbox{end}}(t)\equiv p(\pm L,t).
\end{equation} 
By repeated differentiation of (\ref{ Pend}) with the use of Eqs.\
(\ref{ group}) it can be shown that as $t\to0$ $P_{\mbox{end}}(t) =
O(t^L)$ which means that for sufficiently large boxes the probability
is very close to zero at small $t$.  From the graph of this function
plotted on Fig.\ \ref{FIG1} it is seen that the probability of leaving
the box is practically zero for $t\lesssim0.02L_{\mbox{Box}}^2$.

Let us consider a 1d ``surface'' consisting of $K$ cites with the
cyclic boundary conditions being imposed (site $i=K$ being identical to
site $i=0$).  Let the configuration at time $t$ consists of $n_A$
active atoms, $n_B$ boxed atoms, and $n$ islands.  This configuration
will change with the time-dependent rate (cf.\ Ref.\ \cite{n-fold}
where the only difference is that the rate is constant)
\begin{equation} 
\label{ lambda_t}
\lambda(t)=FK+n_A+n_BP_{\mbox{end}}(t),
\end{equation} 
where the first term describes the rate of deposition of new atoms, the
second corresponds to a hop of an active atom A to a NN site (we remind
that the hopping rate is set to unity) and the last term describes the
rate of B atoms getting out of the boxes.  Because the rate is
time-dependent, we are faced with the necessity to simulate the
nonhomogeneous Poisson process (the EBA is the homogeneous Poisson
process).  We will do this by using the thinning method \cite{thinning}
in its simplest realization with a constant auxiliary rate $\lambda^*$
satisfying
\begin{equation} 
\label{ neq}
\lambda^* \geq \lambda(t).
\end{equation} 
We chose it as
\begin{equation} 
\label{ lambda_star}
\lambda^*=FK+n_A+n_B/L_{\mbox{Box}}.
\end{equation} 
From Fig.\ \ref{FIG1} it is seen that
Eq.\ (\ref{ neq}) is satisfied.

In its most straightforward realization our algorithm 
consists in the following steps.\newline
1. Generate a random
uniform variate $u\in (0,1]$ and advance the time in the boxes as
\begin{equation} 
\label{ t_in_box}
t\to t-\ln(u)/\lambda^*;
\end{equation} 
2. Generate another $u$, calculate the rate 
$\bar{\lambda}=u\lambda^*$, and check whether the inequality 
\begin{equation} \label{ thin}
\bar{\lambda} \leq \lambda(t)
\end{equation} 
holds.  If not, loop back to step 1; if yes go to the next
step;\newline
3a. If $\bar{\lambda} \leq FK$ the deposition event takes
place.  Chose randomly the deposition site and go to step 4;\newline
3b. $ FK<\bar{\lambda} \leq FK+n_A$ corresponds to the atomic jump.
Move a randomly chosen atom to one of NN sites and if this
site is a neighbor to a box or to another atom go to step 4; otherwise
loop back to step 1, diminishing $n_A$ by one if the jump site was a
NN site of an island, so that the atom gets attached to it;\newline
3c. Finally, if $ FK+n_A<\bar{\lambda} \leq \lambda(t)$, an atom 
leaves the box; chose at random the box and the exit 
side; go to the next step;\newline
4. Calculate $\exp(-\epsilon_mt)$ using Eq.\ (\ref{ e_def}) and
find the probability distribution via the FFT in
Eq.\ (\ref{ p_t}).  For each boxed atom generate a discrete random
variable $-L\leq i\leq L$ with the distribution $p(i,t)$ and place
the atom previously in the box centered at $i_B$ at site $(i_B+i \mod
K)$.  Then depending on step 3 nucleate a new island or add the
deposited atom at the random site chosen.  If the site turns out to be
on top of an island move it to the nearest edge, chose it at random if
exactly in the middle.  In this way we avoid the nucleation on tops of
islands.  This prescription is not unique and can be replaced if
necessary; \newline
5. Separate the atoms into groups A and B;
reset the time inside boxes to zero ($t=0$); loop back to step 1.

The majority of the above steps were chosen mainly for their simplicity
with no serious optimization attempted.  In the simulations below the
performance was optimized only through the choice of the box size
$L_{\mbox{Box}}$ which was the same throughout the simulation, though
it seems obvious that by choosing different $L_{\mbox{Box}}$ at
different stages of growth should improve the performance because of
the density which changes with time.  Leaving this and similar
improvements for future studies, in the present paper we
checked the central point of the algorithm which consists in its step
2.  Because with an appropriate choice of $L_{\mbox{Box}}$ most of the
atoms are boxed (up to 100\% at the early stage) and because the
deposition rate $F$ is very small [see Eq.\ (\ref{ R})], at small $t$
the simulation makes a lot of cycles between the 1st and the 2nd steps
due to the small acceptance ratio (see Fig.\ \ref{FIG1}).  Thus, by
simply generating the random variates we simulate diffusion of all
boxed atoms.

\begin{figure} 
\includegraphics[viewport = 0 0 493 304, scale = 0.40]{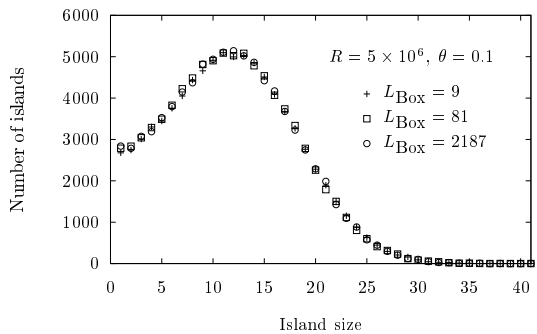}
\caption{\label{FIG2}Illustration of independence of the island size
distribution on the length of the box $L_{\mbox{Box}}$ used in the
simulation algorithm; from 90 to 100\% of atoms were boxed for
$L_{\mbox{Box}}=9$ and from 90 to 100\% were not boxed for
$L_{\mbox{Box}}=2187$ (for further explanations see the text).  The
same statistics corresponding to $10^6$ deposited atoms was gathered
for each box size.}
\end{figure}

\begin{figure} 
\includegraphics[viewport = 0 0 493 301, scale = 0.40]{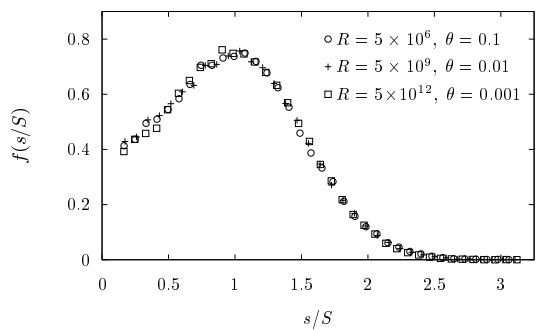}
\caption{\label{FIG3}The scaled island size distribution function
defined in Eq.\ (\ref{ scaling}) as obtained in the kMC simulations
explained in the text. The optimum box sizes were 81, 243, and 729 for
$\theta=0.1$, 0.01, and 0.001, respectively.  Statistics of $5\times
10^5$ atoms was gathered in each of the three cases studied.  Because
of the scaling law $S\propto\theta^{3/4}R^{1/4}$ \cite{1Dscaling} the
number of islands simulated in all three cases was approximately the
same.}
\end{figure}

\begin{figure} 
\includegraphics[viewport = 0 0 314 310, scale = 0.55]{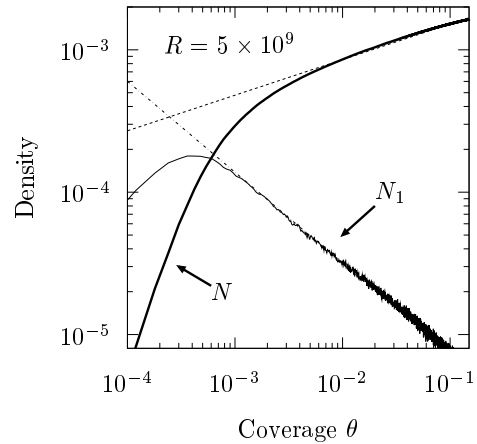}
\caption{\label{FIG4}Island ($N$) and monomer ($N_1$) densities
at different coverages. The dashed line describes the fit to the
asymptotics $N\propto \theta^{1-z}$ with $z=3/4$ \cite{1992,1Dscaling};
the dashed-dotted line is the fit to the asymptotics 
$N_1\propto\theta^{-r}$ with $r\approx0.64$.}
\end{figure}

We simulated the model with the parameters shown in Figs.\
\ref{FIG2}--\ref{FIG4} with $K$ (the system size) in the range
$10^6$--$10^7$ on a 180 MHz MIPS processor.  Our primary goal was to
validate our kMC algorithm and to check the possibility to extend the
parameter ranges achieved in previous studies.  To the best of our
knowledge, we succeeded in carrying over the simulations with the values
of major parameters, such as $R$ and $K$ exceeding those in previous
studies while our smallest value of coverage is the smallest among
those used previously in kMC simulations.  This was achieved with the
maximum execution time (for one run) slightly larger than 2.5 h.  We
expect that with better optimization with modern processors even better
results can be achieved.

Though no systematic study of scaling was attempted,  
the data on the scaling function $f$ defined as \cite{1992}
\begin{equation}
\label{ scaling}
N_s = \frac{\theta}{ S^2}f\left(\frac{s}{S}\right)
\end{equation}
(where $N_s$ is the density of islands of size $s$ and
$S=\sum_{s=2}^{\infty}sN_s$ is the mean island size) presented on
Fig.\ \ref{FIG3} show perfect scaling for all tree cases studied which
differ 6 orders of magnitude in the deposition rate and two orders of
magnitude in coverage.  No dependence of $f(0)$ on $\theta$ found in
Ref.\ \cite{famarescu} is seen in our Fig.\ \ref{FIG3} though the range
of variation of $\theta$ is more than two orders of magnitude larger.
The index $z=3/4$ used in Fig.\ \ref{FIG4} to fit the data on $N\equiv
n/K$ provides better fit then the value $z=1$ suggested in
Ref.\ \cite{famarescu} for the extended islands.  In our opinion, the
point island value is a reasonable choice at very low coverages because
the island sizes became negligible in comparison with the interisland
separations (the gap sizes).  The situation needs further investigation
because another index $r$ was found to be equal to $\sim0.64$ while the
mean field theory predicts it to be 1/2 \cite{1992,1Dscaling}.
Presumably, the value of $R=5\times10^9$ used by us was not
sufficiently large for the scaling to set in.  We note, however, that
it is 500 times larger than that used in Ref.\ \cite{famarescu}.

In conclusion we would like to stress that the technique presented
above can be applied to any separable systems, not only to case
considered in the present paper.  Neither the
availability of an analytical solution is critical.  The solution for
the subsystems can be numerical or even obtained via kMC simulations.
Further modifications may include introduction of several scales,
e.\ g., with the use of the boxes of different sizes as in
Ref.\ \cite{FPT_MC}; the subsystems chosen can be different at
different stages of the simulation.  In brief, we believe that the
technique proposed is sufficiently flexible to allow for the
development of efficient kMC algorithms for broad class of separable
systems.
\begin{acknowledgments}
The authors acknowledge CNRS for support of their collaboration and
CINES for computational facilities. One of the authors (V.I.T.) expresses his
gratitude to University Louis Pasteur de Strasbourg and IPCMS for
their hospitality.
\end{acknowledgments}

\end{document}